\documentclass[12pt]{iopart}
\usepackage{rotating}
\usepackage{graphicx}
\usepackage{latexsym,textcomp}
\usepackage{iopams}
\usepackage{bm}

\newcommand{\PPR}{\ensuremath{\mathcal{P}_{\rm Rot}}}
\newcommand{\PP}{\ensuremath{\mathcal{P}}}

\newcommand{\eqref}[1]{(\ref{#1})}

\begin{document}

\title[Punctured staircase polygons]
{Exact perimeter generating function for a model of punctured staircase polygons \\
}

\author{Iwan Jensen\dag\ and Andrew Rechnitzer\ddag}

\address{
\dag\ Department of Mathematics and Statistics, 
The University of Melbourne, Victoria 3010, Australia \\
\ddag\  Department of Mathematics, University of British Columbia, Vancouver, BC V6T 1Z2, Canada}

\date{\today}

\ead{I.Jensen@ms.unimelb.edu.au,andrewr@math.ubc.ca} 

\begin{abstract}
We have derived the perimeter generating function of a 
 model of punctured staircase polygons in which the internal staircase polygon is
rotated by a 90\textdegree\ angle with respect to the outer staircase polygon. 
In one approach we calculated a long series expansion for the problem and found that 
all the terms in the generating function
can be reproduced from a linear Fuchsian differential equation of order 4. We then
solved this ODE and found a closed form expression for the generating function.
This is a highly unusual and most fortuitous result since ODEs of such high order
very rarely permit a closed form solution. In a second approach we
proved the result for the generating function exactly using combinatorial arguments. 
This latter solution allows many generalisations including to models with other 
types of punctures and to a model with any fixed number of nested rotated staircase
punctures.
\end{abstract}

\submitto{\JPA}


\maketitle

\section{Introduction}

Exactly solved models are special, important and relatively rare in physics.
In many cases real life phenomena are modelled by simplified solvable
models, which despite the simplifications can give us great insight into the 
behaviour of the more complicated fully-fledged problem.  A well-known long 
standing problem in statistical mechanics is to find the perimeter generating 
function for self-avoiding polygons on a  regular two-dimensional lattice.  
Several simplifications of this problem are solvable \cite{Bousquet-Melou96},
but all the simpler models impose an effective directedness or other constraint 
that reduces the problem, in essence, to a one-dimensional problem.  A very 
important and interesting insight gained from these simple models 
(staircase polygons  in particular)   is the conjecture for the limit distribution of area 
moments and scaling function for self-avoiding polygons \cite{Richard01a,Richard02,Richard04}. 

Here we report on the discovery of the exact perimeter generating function for a model of 
punctured staircase polygons. This solution was first conjectured on the basis
of series analysis (see Sections~3 and~4) and subsequently proved
using combinatorial arguments (Section~5). The combinatorial construction admits
many generalisations including to models with other  types of punctures and to a
model with any fixed number of nested rotated staircase punctures (Section~6).

A staircase polygon can be viewed as the intersection of two directed walks starting at the
origin, moving only to the right or up and terminating once the walks join at a vertex.
The perimeter length of a staircase polygon is even. Let us denote by $c_n$ the
number of staircase polygons of perimeter $2n$.  It is well known that 
$c_{n+1}=C_n=\frac{1}{n+1} {2n \choose n}$ are given by  the Catalan numbers $C_n$ and 
that the associated half-perimeter generating function is
\begin{equation}\label{eq:StGf}
P(x) =  \sum_n c_nx^n = \frac12 \left (1-2x-\sqrt{1-4x} \right) \propto (1-\mu x)^{2-\alpha},
\end{equation}
where the connective constant $\mu=4$ and the critical exponent $\alpha=3/2$.
From this it readily follows that  $c_n$ grows asymptotically as
$c_n \sim A_{\rm S} \mu^n n^{\alpha -3}$,
where the critical amplitude $A_{\rm S}=-1/\left (2\Gamma(\alpha-2)
\right)=1/(4\sqrt{\pi})\approx0.141$.

\begin{figure}
\begin{center}
\includegraphics[width=12cm]{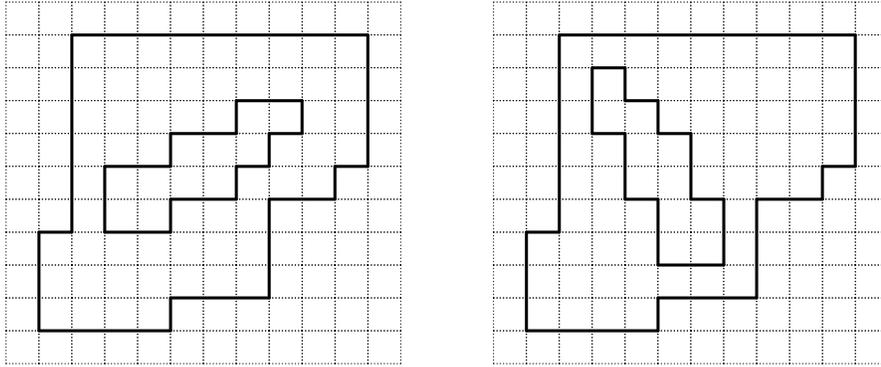}
\end{center}
\caption{\label{fig:poly} The left panel shows a typical punctured staircase polygon and
the right panel a rotated punctured staircase polygon.
}
\end{figure}

Punctured staircase polygons \cite{Guttmann00} are staircase polygons with internal 
holes which are also staircase polygons (the polygons and holes are mutually- as well as
self-avoiding). In a recent paper \cite{Guttmann06a} we studied the problem of
staircase polygons with a single hole and found that the perimeter generating
function can be expressed as the solution of an 8th order linear ODE.
Here we will study the case with a {\em single} hole but where the internal
polygon is rotated by 90\textdegree\ with respect to the main axis. So the
internal staircase polygon is the intersection of two walks starting at a vertex
(the top left-most vertex of the internal polygon) and taking steps only to the
right and {\em down}. We will refer to these objects as rotated punctured
staircase polygons. In Figure~\ref{fig:poly} we have shown an example of each of
the two types of punctured polygons. The perimeter length of
a punctured polygon is the sum of the outer perimeter and the perimeter
of the hole. We denote by $p_n$ the number of punctured staircase polygons 
of perimeter $2n$ and by $r_n$ the number of rotated punctured staircase polygons of
perimeter $2n$. The associated generating functions are $\PP (x)= \sum p_nx^n$
and $\PPR(x)=\sum r_nx^n$, respectively.

It is intuitively clear that since staircase polygons are elongated along the diagonal
of growth there are many more restrictions on the placement of the rotated polygon
and hence we expect $r_n \leq p_n$. The difference between the two cases can be
made more explicit by noting that any polygon has a {\em minimal bounding rectangle}
which is the smallest rectangle completely containing the polygon. In the rotated case 
the outer polygon is excluded from the minimal bounding rectangle of the hole while
in the aligned case the outer polygon may enter the minimal bounding rectangle
of the hole (as is the case in the left panel of Figure~\ref{fig:poly}).

In \cite{Guttmann00} it was proved that the connective constant $\mu$ of $k$-punctured 
polygons (polygons with $k$ holes) is the same as the connective constant of unpunctured polygons. 
Numerical evidence clearly indicated that the critical exponent $\alpha$ increased by $3/2$ per 
puncture (this was proved for a single puncture and conjectured in the general case).  
In recent work Richard, Jensen and Guttmann \cite[Theorem 2]{Richard07}  proved the exponent
formula for a finite number of punctures and proved an exact formula for the leading 
amplitude of punctured staircase polygons. It it worth noting that
the proofs in these papers never considered restrictions on the placement of the
internal polygon, that is, the hole could be placed in any way one pleases. The
results,  therefore, carry over unaltered to the problem of rotated punctured
staircase polygons. Interestingly, this means that the leading asymptotic forms
of $r_n$ and $p_n$ are exactly the same; any differences arise only from
sub-dominant correction terms. In particular the dominant singular behaviour of
$\PP (x)$ and $\PPR(x)$ is a simple pole at $x=x_c=1/4$ and thus $p_n \sim
A_{\rm P} 4^n$ and $r_n \sim A_{\rm R} 4^n$, where we expect that $A_{\rm P}
=A_{\rm R} $.

The rest of the paper is organised as follows. In Section~\ref{sec:enum} we
briefly describe the algorithm used to count the number of punctured polygons.
Section~\ref{sec:fde}  describes how we found the underlying ODE and gives a
brief review of its critical properties. Then, in Section~\ref{sec:sol},
we give the closed form solution for the generating function.
Section~\ref{sec:comb} contains an outline of the rigorous combinatorial
derivation of the generating function while Section~\ref{sec:gen} contains a
brief outline of results obtained by generalising the combinatorial formula.
Finally, in Section~\ref{sec:conc} we briefly discuss our findings.

\section{Computer enumeration \label{sec:enum}}

Rotated punctured staircase polygons were counted using a generic algorithm
designed to count punctured convex polygons, which in turn was based on
an algorithm for counting general self-avoiding polygons \cite{IGE80}, but 
with restrictions so as to limit the count to various types of convex polygons. 
A  polygon is convex if the perimeter is equal to that of its minimal bounding
rectangle (the smallest rectangle into which one can fit the polygon).
A staircase polygon is  a restricted convex polygon and is produced
by demanding that two diagonally opposite corners of the minimal
bounding rectangle are part of the convex polygon. A standard punctured
staircase polygon is then produced by demanding that the outer and inner
 polygons must include, say, the lower left and upper right corners of their respective 
minimal bounding rectangles.  For rotated punctured staircase polygons
the inner polygon must include the upper left and lower right corners as illustrated
in Figure~\ref{fig:poly}. The algorithm used to count punctured staircase polygons
(the aligned case) is described in \cite{Guttmann06a,Guttmann06}.

\Table{\label{tab:PPcoef}
The number $r_n$  of rotated punctured staircase polygons.}
\br
$n$ & $r_n$ & $n$ & $r_n$ & $n$ & $r_n$  \\
\mr
8   & 1  & 19 & 201099320  & 30 & 1761048979430768 \\
9   & 12 & 20 & 889594210  & 31 & 7344148372848448 \\
10 & 94 & 21 & 3896177956 & 32 & 30553399525543917 \\ 
11 & 604 & 22 & 16920602244 & 33 & 126830292729207600 \\
12 & 3461 & 23 & 72954802376  & 34 & 525432597401411262 \\ 
13 & 18412 & 24 & 312595497011 & 35 & 2172784129140676636 \\ 
14 & 93016 & 25 & 1332153819572 & 36 & 8969907982862433143 \\
15 & 452500 & 26 & 5650155211024 & 37 & 36973458557889104804 \\ 
16 & 2139230 & 27 & 23864065957572 & 38 & 152186202561672129880 \\
17 & 9890404 & 28 & 100418115489408 & 39 & 625590993787945461804 \\
18 & 44921002 & 29 & 421151065542880  & 40 & 2568489385305061560252 \\
\br
\endTable

Using this algorithm we calculated $r_n$ up to half-perimeter $n=125$. Since the
smallest punctured polygon has half-perimeter $8$ this gives us a 118 non-zero
terms. In Table~\ref{tab:PPcoef} we have listed the first few values of $r_n$.
For comparison we mention that the first discrepancy with $p_n$ is
at $n=12$ and $p_{12}-r_{12}=2$. Also note, for comparison, that
$p_{40}/r_{40}\approx 1.041$.

\section{The Fuchsian differential equation \label{sec:fde}}

Recently Zenine {\it et al} \cite{Zenine04,Zenine05,Zenine05a} used
series analysis to obtain linear ODEs satisfied by the 3-
and 4-particle contributions  $\chi^{(3)}$ and $\chi^{(4)}$ to the Ising model
susceptibility. In \cite{Guttmann06a,Guttmann06} we
used their method to find linear ODEs for the perimeter generating functions of
punctured staircase polygons and three-choice polygons. We have used this
technique to find a linear ODE satisfied by the generating function
 $\PPR (x)$ for rotated punctured staircase polygons. 
We briefly outline the method here. Starting from a (long) series 
expansion for a function $G (x)$ we look for a linear differential equation of order $m$ 
of the form

\begin{equation}\label{eq:de}
\sum_{k=0}^m P_k(x) \frac{\rmd^k}{\rmd x^k}G(x) = 0,
\end{equation}
such that $G(x)$ is a solution to this homogeneous linear differential equation, where
the $P_k(x)$ are polynomials. In order to make it as simple as possible we start
by searching for a Fuchsian  \cite{Ince27} equation. Such equations have only regular singular points. 
Computationally the Fuchsian assumption simplifies the search for a  solution.
From the general theory of Fuchsian  \cite{Ince27} equations it follows 
that the degree of $P_k(x)$ is at most $n-m+k$ where $n$ is the degree of $P_m(x)$. 
To simplify matters further (reduce the order of the unknown polynomials) it is advantageous
to explicitly assume that the origin and $x=x_c=1/4$ are regular singular points
and to set $P_k(x)=Q_k(x)S(x)^k$, where $S(x)=x(1-4x)$. Thus when searching for a solution 
of Fuchsian type there are only two parameters, namely the order $m$ of the differential 
equation and the degree $q_m$ of the polynomial $Q_m(x)$. Since the degree
of the imposed factor $S(x)$ is 2 the restriction on the degree of $P_k(x)$ means that the 
degree of $Q_k(x)$ is at most $q_m+m-k$. The number of unknown coefficients is thus
$L=\sum_{j=0}^m ( q_m+j+1)-1=(m+1)q_m+(m+2)(m+1)/2-1$, where we get one less
unknown by demanding that the leading coefficient in $Q_m(x)$ is~1.
   
We then search systematically for solutions by varying $m$ and $q_m$. In this
way we found a solution with $m=4$ and $q_m=4$, which required the determination
of only $L=34$ unknown coefficients. We have 118 non-zero terms in the half-perimeter 
series and thus have 84 additional terms with which to check the correctness of our solution.
This should be contrasted with punctured staircase polygons \cite{Guttmann06a} where
we first found a solution with $m=10$ and $q_m=11$, which required the determination
of $L=186$ unknown coefficients.  The lowest order ODE we found had order $m=8$ with
 $q_m=27$, which requires the determination of $287$ unknown coefficients. 
 So clearly the restrictions imposed by the rotation of the internal polygon results
 in a much simpler problem ({\it a priori} there was obviously no reason to believe this would
 be the case).

The (half)-perimeter generating function $\PPR (x)$ for rotated punctured staircase polygons
is a solution to the linear differential equation of order 4

\begin{equation}
\sum_{k=0}^4 P_n(x) \frac{\rmd^k}{\rmd x^k}\PPR (x) = 0
\label{eq:PPfde}
\end{equation}
with

\begin{eqnarray}
\fl
P_4(x)=x^2\,{\left( 1\! -\! 4\,x \right) }^4\,\left( 3 \!+\! 4\,x \right)
\,\left( -252 \!+\! 300\,x \!+\! 2365\,x^2 \!+\! 1800\,x^3 \right),  \nonumber
\\
\fl
P_3(x)=x(1\!-\!4x)^3\left (5292 \!+\! 1872\,x \!-\! 56127\,x^2 \!-\! 115700\,x^3
\!-\! 97280\,x^4 \!-\! 28800\,x^5 \right),  \nonumber \\
\fl
P_2(x) = 24(1\!-\!4x)^2 \left(  -378 \!-\! 189\,x \!+\! 5565\,x^2 \!+\!
4085\,x^3 \!+\! 1480\,x^4 \!+\! 4090\,x^5 \!+\! 3600\,x^6 \right ), \\
\fl
P_1(x)=24(1\!-\!4x)\left(  -126 \!-\! 1113\,x \!+\! 1250\,x^2 \!+\! 24540\,x^3
\!-\! 2805\,x^4 \!-\! 44960\,x^5 \!-\! 28800\,x^6 \right), \nonumber  \\
\fl
P_0(x) = 24\left(  504 \!+\! 672\,x \!-\! 12200\,x^2 \!-\! 38475\,x^3 \!+\!
112600\,x^4 \!+\! 228800\,x^5 \!+\! 115200\,x^6 \right). \nonumber
\end{eqnarray}

The singular points of the differential equation are given by the roots of $P_4(x)$.
One can easily check that all the singularities (including $x=\infty$) are
{\em regular singular points} so equation~\eqref{eq:PPfde} is indeed of the
Fuchsian type. It is thus possible using the method of Frobenius to obtain from
the indicial equation the critical exponents at the singular points. These are
listed in Table~\ref{tab:PPexp}.

\Table{\label{tab:PPexp}
Critical exponents for the regular singular points of the Fuchsian differential
equation satisfied by $\PPR (x)$.}
\br
Singularity & \centre{2}{Exponents} \\
\mr
$x=0$ & $-8$ &$-7$ &  $ -4$ &  $0$ \\
$x=1/4$ & $-1$ & $-3/4$ & $-1/2$ & $-1/4$ \\
$x=-3/4$ & $0$ & $1/2$ & $1$  & $2$ \\
$1/x=0$ & $6$  & $ 3/2$ & $ 7$ & $15/2$ \\
$Q_4(x)=0$ & $0$ & $1$ & $ 2$ & $ 3$ \\
\br
\endTable

We shall now consider the local solutions to the differential equation around each singularity. 
Recall that in general it is known \cite{Forsyth02,Ince27} that if the indicial equation yields $k$ 
critical exponents which differ by an integer, then the local solutions {\em may} 
contain logarithmic terms up to $\log^{k-1}$. However, we have found in the analysis of
previous problems that for the Fuchsian equations of the type
described by equation~\eqref{eq:PPfde} {\em only} multiple roots of the indicial
equation give rise to logarithmic terms in the local
solution around a given singularity, so that a root of multiplicity $k$  gives
rise to logarithmic terms up to $\log^{k-1}$.  In particular this means that
near any of the 3 roots of $Q_4(x)=-252 \!+\! 300\,x \!+\! 2365\,x^2 \!+\!
1800\,x^3 $, the local solutions have no logarithmic terms and the solutions are
thus {\em analytic} since all the exponents are positive integers. The roots of
$Q_4(x)$ are thus {\em apparent singularities}
\cite{Ince27,Forsyth02} of the Fuchsian equation~\eqref{eq:PPfde}.  This will
become completely self-evident in the next section where we present a closed
form solution of $\PPR(x)$. So the points of interest are the physical critical
point $x=x_c=1/4$, where the dominant singularity is a simple pole, modified by
3 correction terms with exponents that increase in steps of $1/4$. At the
non-physical critical point $x=-3/4$ the function has a simple square
root singularity.

\section{The solution to the ODE \label{sec:sol}}

Given an ODE it is often useful to look for simple solutions. In this case we first looked for
solutions of the form $P(x)/(1-4x)^{\gamma}$, where $P(x)$ is a polynomial and
$\gamma= 1, 3/4, 1/2$  or $1/4$.
In this fashion we discovered the solutions $F_1(x)$ and $F_2(x)$ listed below.
This gave us great hope that we could find a solution to the full problem since the
two simple solutions can be used to write the ODE  as a product of three differential
operators of order 2, 1 and 1, respectively. As it turned out the {\tt dsolve} package in Maple was 
up to the task and readily found four solutions, including $F_1(x)$ and $F_2(x)$.
The remaining two solutions as supplied by {\tt dsolve}  were quite complicated
expressions, but with a bit of work we managed to simplify (largely by hand) to
the expressions $F_3(x)$ and $F_4(x)$ given below.

The four linearly independent solutions to the ODE are:

\numparts
\begin{eqnarray}
F_1(x) ={\frac {1-8\,x+16\,{x}^{2}-4\,{x}^{3}}{1-4\,x}}, \\
F_2(x) ={\frac {1-6\,x+6\,{x}^{2}}{\sqrt {1-4\,x}}}, \\
F_3(x) =\frac{1}{\sqrt{2}}\,{\frac {\sqrt {2+\sqrt {3+4\,x}} \left( 3-8\,x+2\,{x}^{2}-\sqrt {
3+4\,x} \left( 1-2\,x \right)  \right) }{ \left( 1-4\,x\right) ^{3/4}}}, \\
F_4(x) =\frac{1}{\sqrt{2}}{\frac { \left( 3-8\,x+2\,{x}^{2}+\sqrt {3+4\,x} \left( 1-2\,x
 \right)  \right) }{\left (1-4\,x \right )^{1/4}\sqrt {2+\sqrt {3+4\,x}}
}},
\end{eqnarray}
\endnumparts
and the  generating function is simply
\begin{equation}
\label{eqn ode sol}
\PPR(x) = -\frac14 \left [ F_1(x)-F_2(x)+F_3(x)-F_4(x) \right ];
\end{equation}
we prove this formula in the next section. Obviously, $\PPR(x)$ is dominated
asymptotically by $-\frac14 F_1(x)$, so the leading amplitude $A_{\rm
R}=-\frac14 \left [ (1-4x)F_1(x) \right ] _{x=1/4}/\Gamma(1)=1/64$,
which is exactly the same as the leading amplitude of punctured staircase
polygons and equal to the proved formula  \cite{Richard07}.

\section{Combinatorial construction of  a staircase polygon with a rotated
staircase hole
\label{sec:comb}}

Before we consider the construction, it is important to note is that
because the inner polygon (the hole) is rotated $90^\circ$ from the outer
polygon, it follows that the outer polygon must avoid not only the inner
polygon, but also its minimal bounding rectangle. This considerably simplifies
the problem.

\begin{figure}
\begin{center}
\includegraphics[width=12cm]{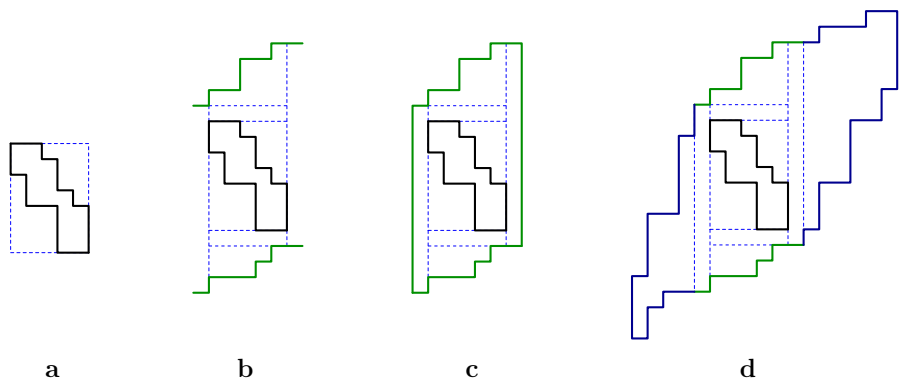}
\end{center}
\caption{\label{fig:cons}  
The four stages in the combinatorial construction of a rotated staircase polygon.
We start (a) from a staircase polygon sitting in it's minimal bounding rectangle.
Next (b) we attach directed walks above and below the rectangle and (c) connect
the walks to form an outer polygon. Finally (d) we complete the construction by attaching
staircase factors on the left and right sides of the outer polygon. }
\end{figure}

We build the punctured polygon in four steps (see Figure~\ref{fig:cons}).
\begin{enumerate}
\item[{\bf a:}] Construct the inner polygon, enumerating it by total perimeter,
  width and height. A wasp-waist decomposition or a column-by-column
  construction gives the following generating function \cite{Bousquet-Melou96}:
  \begin{eqnarray}
    P(w,h;x,y) = x^2y^2 w h + (x^2w+y^2h) P(w,h;x,y) + P(w,h;x,y)^2 \nonumber  \\
    = \frac{1}{2} \left( 1\! -\! x^2w\!-\!y^2h 
    - \sqrt{1\!-\!2x^2w\!-\!2y^2h\! +\! x^4w^2\!-\!x^2y^2wh\!+\!y^4h^2} \right)
    \label{eq:CStair}
  \end{eqnarray}
  where $x,y$ are conjugate to the number of horizontal and vertical bonds and
  $w, h$ are conjugate to the width and height of the associated minimum
  bounding rectangle.

\item[{\bf b:}]  We then attach two directed paths to the bounding rectangle of
  the polygon; one above and one below. If the polygon has width $w$,
  then each of these paths must start and end with a horizontal bond, contain
  a total of $w+2$ horizontal bonds.  The generating function of a directed path is 
  \begin{equation}
    D(x,y) = \frac{1}{1-x-y} = \sum_n \frac{x^n}{(1-y)^{n+1}}.
  \end{equation}
  Hence the generating function of directed walks with exactly  $w$ horizontal
bonds is $x^w/(1-y)^{w+1}$. So if we wish to attach a directed path of width
$w+2$ above a staircase polygon of width $w$, then the corresponding generating
function is $P(wx/(1-y), h;x,y)\cdot x^2 / (1-y)$. If we attach two
directed paths  (one above and one below) we obtain $P(wx^2/(1-y)^2,h;x,y)
\cdot x^4 / (1-y)^2$.
     
Next we have to insert a (positive) number of rows between the
directed paths and the bounding rectangle. We also need to keep track of some
extra variables. In particular we need to know the height of the leftmost and
rightmost columns (the distances between the end-points of the directed walks). 
Let these be counted by the variables $s$ and $t$, respectively. 
   
Inserting a positive number of rows between the directed paths and the
bounding rectangle is equivalent to multiplying by
$\left(\frac{st}{1-st}\right)^2$. Since each directed path adds to either the
height of the leftmost or rightmost column we need to modify what we did
above. In particular the generating function of the upper directed path becomes
$(1\!-\!x\!-\!yt)^{-1}$ and that of the lower path becomes $(1\!-\!x\!-\!ys)^{-1}$. Thus we
arrive at
\begin{equation}
    \fl
     Q(s,t;x,y) = \left(\frac{st}{1-st}\right)^2 \frac{x^4}{(1-sy)(1-ty)}
    P\left(\frac{x^2}{(1-sy)(1-ty)},st,x,y \right)
\end{equation} 
where $Q$ is the generating function of open configurations as drawn in
Figure~\ref{fig:cons}(b).

\item[{\bf c:}]  I the next step we simply add vertical bonds to each of the open 
ends thus connecting the directed paths  so as to form a closed object. This
corresponds to mapping $s \mapsto sy$ and $t \mapsto ty$.
Hence our generating function is
\begin{eqnarray}
     R(s,t;x,y) = Q(sy,ty;x,y) \nonumber \\
    = \left(\frac{sty^2}{1\!-\!sty^2}\right)^2 \frac{x^4}{(1\!-\!sy^2)(1\!-\!ty^2)}
    P\left(\frac{x^2}{(1\!-\!sy^2)(1\!-\!ty^2)},sty^2,x,y \right)
\end{eqnarray} 
which counts the closed configurations as drawn in Figure~\ref{fig:cons}(c).

\item[{\bf d:}]  The punctured staircase polygons are completed by adding left
and right staircase polygon factors to each side. Roughly speaking, this is
done by substituting $s \mapsto \sigma(x,y)$ and $t \mapsto \sigma(x,y)$ where
$\sigma$ is very nearly the staircase polygon generating function.

More specifically we do a Temperley/Bousquet-M\'elou-like column-by-column
construction to obtain the appropriate generating function. 
In \cite{Bousquet-Melou96} the functional equation used to
enumerate staircase polygons by perimeter and area is given by
\begin{equation}
     F(s) = A(s) + \frac{x^2s}{(1-s)(1-sy^2)} \left(F(1) - F(s) \right)
\end{equation}
where $A(s)$ is the ``\emph{seed}'' configuration to which we append columns
and $F(s)$ is the generating function of the resulting staircase polygons.

This functional equation can be solved using the kernel method. First take
all the $F(s)$ terms to the left-hand side of the equation:
\begin{equation}
     F(s) \left(1 + \frac{x^2s}{(1-s)(1-sy^2)} \right) = A(s) + \frac{x^2s}{(1-s)(1-sy^2)} F(1) 
\end{equation}
Now set $s$ to a value that takes the coefficient of $F(s)$ (the kernel) to
be zero. This value is
\begin{equation}
    \sigma(x,y) = \frac{1}{2y^2} 
    \left( 1\! -\! x^2\!+\!y^2 - \sqrt{1\!-\!2x^2\!-\!2y^2\! -\!2x^2y^2\!+\! x^4\!+\!y^4} \right),
\end{equation}
which is very nearly the staircase polygon generating function. Setting $s$
to this value leaves
\begin{equation}
     0  = A(\sigma) + \frac{x^2 \sigma}{(1-\sigma)(1-\sigma y^2)} F(1)= A(\sigma) - F(1)
\end{equation}
where we have made use of the fact that $x^2 \sigma = - (1-\sigma)(1-\sigma y^2)$.  
We do not care about $F(s)$ and only want $F(1)$ (\emph{i.e.} we
do not need to know the leftmost and rightmost column heights of our punctured
polygons) and we are done.
    
\end{enumerate}

The generating function of staircase polygons with a rotated staircase hole
is given by
\begin{equation}
    \fl
     S(x,y) = R(\sigma,\sigma;x,y)
   = \left(\frac{\sigma^2 y^2}{1-\sigma^2y^2}\right)^2  \frac{x^4}{(1-\sigma y^2)^2}
    P\left(\frac{x^2}{(1-\sigma y^2)^2},\sigma^2 y^2,x,y \right)
\end{equation}
Of course, the punctured polygons always have an even number of horizontal
and vertical bonds,  so we can replace $x \mapsto \sqrt{x}$ and $y \mapsto
\sqrt{y}$ to obtain the half-perimeter generating function:
\begin{equation}\label{eq:Cfinal}
S(x,y) = \left(\frac{\sigma^2 y}{1-\sigma^2y}\right)^2
\frac{x^2}{(1-\sigma y)^2}
P\left(\frac{x}{(1-\sigma y)^2},\sigma^2 y,x,y \right)
\end{equation}
where $P$ is the solution of $P = xywh + (xw+yh)P + P^2$ and $\sigma$ is the
solution of $(1-\sigma)(1-y\sigma)+x\sigma=0$. One may verify that setting
$y=x$ does indeed recover the generating function given in
equation~\eqref{eqn ode sol}

\section{Generalisations of the combinatorial results \label{sec:gen}}

Here we shall consider three generalisations of the combinatorial result. In the
first generalisation we look at the case where we do not count the perimeter of 
the hole (in other words the perimeter of the punctured polygon is equal to the
 perimeter of the external polygon only). In the second generalisation we look at 
 what happens if we replace the internal polygon by a convex polygon of a different 
 type. In the third generalisation we show that the formula~\eqref{eq:Cfinal}
can be used to derive the generating function for nested staircase polygons.
 
 The first generalisation is simply obtained by noting that in
formula~\eqref{eq:Cfinal} the variables $x$ and $y$ `counting' the perimeter can
be separated in order to  count the external and internal perimeter. In fact
going back through the derivation  we note that in equation~\eqref{eq:Cfinal}
only the last (right-most) occurrence of $x$ and $y$  come from the internal
polygon (all the other occurrences arise from the outer polygon).
 We thus find that the generating function for rotated punctured staircase
polygons counted only by {\em external} perimeter is
    \begin{equation}\label{eq:RotExt}
     S(x,y)   = \left(\frac{\sigma^2 y}{1-\sigma^2y}\right)^2 
    \frac{x^2}{(1-\sigma y)^2}
    P\left(\frac{x}{(1-\sigma y)^2},\sigma^2 y,1,1 \right) 
    \end{equation}
Analysis of this solution shows that again it is the solution to a 4th order ODE and
there is a major change in the critical behaviour. The function now has a square
root singularity at $x=4/25$, that is a behaviour $\propto \sqrt{4-25x}$, so the
connective constant changes from 4 to 25/4. There is still a singularity at
$x=1/4$, and interestingly the behaviour around it is quite complicated
with exponents $-1$, $-1/2$, 0 and 1/2.

The second generalisation is obtained by noting that in the derivation of 
formula~\eqref{eq:Cfinal} we never really used the fact that the internal
polygon  was a staircase polygon we only relied on the fact that the external
polygon is excluded from the minimal bounding rectangle of the hole. This means
that we can replace $P(w,h,x,y)$ with the generating function of another type
of polygon provided we know how to enumerate them by width, height  and
perimeter. In Table~\ref{tab:crit} we summarise the results for the critical
behaviour at the leading singularity when the puncture is a convex polygon.
Firstly we note that  the dominant singularity remains at $x=1/4$. However, the
critical exponent changes depending on the asymptotic growth of the generating
function of the puncture. When the asymptotic growth of the puncture is slower
than or equal to that of staircase polygons we have a simple pole at $x=1/4$.
However, in the case of directed convex and the case of convex polygons (both of
which are asymptotically more numerous than staircase polygons) the critical
exponent changes to reflect a faster growth in the number of the associated
punctured polygons.
 
\begin{table}
\caption{\label{tab:crit} The critical point and exponents of staircase polygons
with a puncture which is a convex polygon. The left-most columns give the
behaviour of the convex polygon used for the puncture while the right-most columns 
is the behaviour once `wrapped' in an outer staircase polygon.}
\begin{indented}
\item[]\begin{tabular}{@{}lll|ll}
\br
Puncture type& $x_c$ & Exponent & $x_c$ & Exponent \\
\mr
Rectangle & 1 & $-2$ & 1/4& $-1,  -1/2$ \\
Ferrers & 1/2 & $-1$ & 1/4 & $-1,  -1/2$ \\
Stack & $(3\!-\!\sqrt{5})/2$ & $-1$ & 1/4 & $-1,  -1/2$ \\
Staircase & 1/4 & 1/2 & 1/4  & $-1,  -3/4,  -1/2,  -1/4$\\
Directed & 1/4 & $-1/2$ & 1/4 & $-5/4,  -3/4$\\
Convex & 1/4 & $-2,  -3/2$ & 1/4 & $-2,  -7/4,  -3/2,  -5/4$ \\
\br
\end{tabular}
\end{indented}
\end{table}
 
The third generalisation is obtained by noting that formula~\eqref{eq:Cfinal} 
can be iterated. This gives us (at least in principle) solutions for any fixed
number of staircase polygons nested within one another. Where at each level the
outer polygon avoids the minimal bounding rectangle of the inner  polygon(s).
In order to prove the iteration procedure we note the following facts:

\begin{enumerate}
\item As already noted above the  $x$ and $y$ variables can easily be distinguished
so as to count the outer and inner perimeters separately. 
\item We don't really need to keep track of the internal perimeter in detail.
\item Because of the convexity constraint the $(x,y)$-variables from the outer
polygon automatically counts the width and height so
the $w$ and $h$ variables in the original $P$ are not really needed. 
\item We have no restriction on what we put inside the hole.
In particular we could put a staircase polygon with a rotated staircase polygon
into the hole.
\end{enumerate}

With this in mind we can see formula~\eqref{eq:Cfinal} as an operator on
generating functions along the lines of ``wrap me in a staircase polygons but
don't enter my minimal bounding rectangle",

\begin{equation}\label{eq:wrapop}
T[P(x,y)] \mapsto  B(x,y)P(xz/(1\!-\!\sigma y)^2,\sigma^2 yz),
\end{equation}
where $B(x,y)$ is the prefactor $(\sigma^2 y/(1\!-\!\sigma^2 y))^2 x^2/(1\!-\!\sigma y)^2$.
By iteration we then find that the generating function for $k$-nested rotated staircase 
polygons is
\begin{equation}\label{eq:Pnest}
\PP_k(x,y) = T[\PP_{k-1}(x,y)] , \mbox{ with } \PP_0(x,y)= P(x,y).
\end{equation}

The solution for the twice nested case $\PP_2(z) = \PP_2(z,z)$ has been confirmed by 
enumeration results. The formula~\eqref{eq:Pnest} gives us (at least in
principle) solutions for any fixed number of staircase polygons nested within
one another. In reality the solutions are pretty nasty and so far we have
determined mainly the leading asymptotic behaviour. In summary we find that the
generating functions $\PP_k(z)=\PP_k(z,z)$ have a singularity
at $z=1/4$ with leading exponent $-2+1/2^{k-1}$. We have confirmed the exponent
value for $k$ up to 10 by doing formal expansions around $z=1/4$.
It appears that the singularity is of order 
$2^{k+1}$ and that the full set of exponents is given by $-2+1/2^{k-1} + j/2^{k+1}$, $j=0,\ldots, 2^{k+1}-1$.
We have already seen that this behaviour is true for $\PP_1(z)$  from the closed formed solution.
For $k=2$ we took the formula for $\PP_2(z)$ and expanded in a series to order 1000.
Using the method described in Section~\ref{sec:fde} we
managed to find the exact ODE which is of order 8 as expected  with
degree of the leading polynomial equal to  66. Solving the indicial equation
confirms that the exponents at $z=1/4$ are $-12/8$, $-11/8$, $-10/8$, $-9/8$, $-8/8$, $-7/8$, 
$-6/8$, and $-5/8$ in complete agreement with our conjecture.

\section{Summary and Discussion \label{sec:conc}}

Using series expansions for rotated punctured staircase polygons we found that
the half-perimeter generating function $\PPR(x)$ satisfies a fourth order Fuchsian ODE.
We then solved this ODE and found a closed form solution for  $\PPR(x)$. The solution
is dominated by a function $-F_1(x)/4$ with a simple root at $x=x_c=1/4$. There are three
sub-dominant correction terms $-F_3(x)/4$ which has critical exponent $-3/4$,
$F_2(x)/4$ with exponent $-1/2$ and $F_4(x)/4$ with exponent $-1/4$. In addition there
is a square root singularity at $x=-3/4$. 

This should be compared to our analysis \cite{Guttmann06a} of the ODE satisfied by the 
generating function $\PP(x)$ for punctured staircase polygons, which showed that near 
the physical critical point $x=x_c=1/4$ 
\begin{equation}\label{eq:xc}
\PP(x) \sim  \frac{A(x)}{(1-4x)}  + \frac{B(x) + C(x) \log(1-4x)}{\sqrt{1-4x}},
\end{equation}
where $A(x)$, $B(x)$ and $C(x)$ are analytic in a neighbourhood of $x_c$. 
In addition $\PP(x)$  has a singularity on the negative $x$-axis, at $x=x_-=-1/4$ with the singular behaviour
\begin{equation}\label{eq:xm}
\PP(x) \sim D(x) (1+4x)^{13/2},
\end{equation}
where again $D(x)$ is analytic near $x_-$.  Finally, the ODE also had a pair of 
singularities at $x=\pm \rmi/2$ and at the
roots of $1+x+7x^2$ (see \cite{Guttmann06a}  for further details), but since these
singularities lie in the complex plane outside the physical disc $|x| \leq x_c$ 
their contributions are exponentially suppressed  asymptotically. 

We argued in the introduction and demonstrated in Section~\ref{sec:sol} that
$p_n$ and $r_n$ have exactly the same asymptotic form to leading order.
Any differences between the two problems only appear in the sub-dominant
correction terms. The dominant correction
term for $\PP(x)$ is $\propto  \log(1-4x)/\sqrt{1-4x}$, which is weaker than 
the first correction term $-F_3(x)/4\propto (1-4x)^{-3/4}$ for $\PPR(x)$.
 The amplitudes of both these correction
terms are negative, namely,  $-3\sqrt{3}/(32\pi^{3/2})\approx-0.029$ and
$-\frac14\left [(1-4x)^{3/4} F_3(x)\right ]_{x=x_c}/\Gamma(3/4) = 
-1/\left (16\sqrt{2}\Gamma(3/4)\right ) \approx -0.036$, respectively.
So this bears out our intuition that $r_n \leq p_n$, since the placement of a
rotated inner staircase polygon faces  more restrictions than  the placement of 
aligned inner and outer staircase polygons. These differences  indicate that
combinatorial arguments for a proof of sub-dominant behaviour must be quite
subtle!

The combinatorial derivation of the expression for $\PPR(x)$ turned out to be
very interesting. In particular it allowed us to prove exact results for several 
generalisations of the original model. In the case where we only count the
external perimeter we find a change in the value of dominant singularity
to $x_c=4/25$. There is still a singularity at $x=1/4$ and interestingly
the behaviour at this (now sub-dominant) critical point is much more complicated
than at the leading singularity. We also studied the case where the puncture
was replaced by a different type of convex polygon (with the external staircase
polygon avoiding its minimal bounding rectangle). The major finding of interest
was that the dominant singularity remained at $x=x_c=1/4$. However, the critical
exponent changed depending on the asymptotic growth of the generating function
of the puncture. When the asymptotic growth of the puncture is slower than or equal
to that of staircase polygons we have a simple pole at $x=x_c=1/4$. However, in the
case of directed convex and the case of convex polygons (both of which are 
asymptotically more numerous than staircase polygons) the critical exponent changes
to reflect a faster growth in the number of the associated punctured polygons. 
Finally, we looked at the case of nested staircase polygons in which case we
can derive formulas for a fixed number $k$ of nested staircase polygons. We found
that the singularity remains at $x=1/4$ with a leading exponent $-2+1/2^{k-1}$;
while we have observed this for $1\leq k \leq 10$,  we have not proved it. 

Another interesting open question is the behaviour of staircase polygons with an
arbitrary number of nested polygons (that is the function formed by summing
$\PP_k(x)$). This work can also be extended to different types of outer
polygons. The method of Section~\ref{sec:comb} should generalise to other types
of convex polygons. Also it might be interesting to look at counts by area as
well.

\section*{E-mail or WWW retrieval of series}

The series for the generating function studied in this paper 
can be obtained via e-mail by sending a request to 
I.Jensen@ms.unimelb.edu.au or via the world wide web on the URL
http://www.ms.unimelb.edu.au/\~{ }iwan/ by following the instructions.

\ack

We would like to thank M Zabrocki for his comments on the combinatorial constructions 
in Section 5 and C Richard and A J Guttmann for their useful comments on the manuscript.
IJ gratefully acknowledge financial support from the Australian Research Council.
AR gratefully acknowledges financial support from NSERC Canada.

\end{document}